\documentclass[12pt,preprint]{aastex}
\usepackage{float}
\usepackage{graphicx}
\usepackage{epstopdf}
\usepackage{txfonts}
\usepackage{natbib}
\usepackage{color}
\usepackage[normalem]{ulem}
\usepackage{subfig}
\usepackage{CJK}
\begin{document}

\title{Correlation Analysis for $\gamma$-ray  and Broad Line Emissions of Fermi Blazars}
\author{ L. X. Zhang, J. H. Fan, Y.H. Yuan}
\altaffiltext{1}{Center for Astrophysics, Guangzhou University, Guangzhou 510006, China}
\altaffiltext{2}{Astronomy Science and Technology Research Laboratory of Department of Education of Guangdong Province, Guangzhou 510006,China}
\begin{abstract}
{In a standard model of active galactic nuclei (AGNs),
 there is a supermassively central black hole
surrounded by an accretion disk with the jet coming out perpendicularly to the disk plane.
Theoretical works suggest that there is a connection between the jet and the accretion disk.
To investigate such a connection,
 people use the correlation between the radio emission ( or $\gamma$-ray emission) and the broad line emission.
However,
 it is well known that the radio (or $\gamma$-ray ) emission is strongly beamed in blazars.
In this sense, we should consider the beaming effect when we discuss the jet--accretion disk connection.

In this work, we compiled a sample of 202 Fermi/LAT blazars  with available broad
 line emissions. Out of the 202 sources, 66  have known Doppler factors.
The correlation between $\gamma$-ray and broad-line emission,
and that between radio and broad-line emission  are investigated by removing the
effects of redshift and beaming boosting for the whole sample and the subclasses,
flat spectrum radio quasars (FSRQs) and BL Lacertae objects (BL Lacs) respectively.
 We obtained a strong positive correlation between $\gamma$-ray and broad-line emission
and between radio and broad-line emission for the 202 blazars;
It's worth noting that the correlation still exists after removing redshift effect.
 For the 66 sources with Doppler factors,
 there is also a positive correlation between $\gamma$-rays and broad-line emission after removing the Doppler factors,
as well as that between radio and broad-line emission.

Our analysis suggest that
1. There are  strong correlations between the $\gamma$-ray and the broad line emission for the whole blazar sample and their subclasses.
The correlations exist when the redshift effect is removed for the whole sample and their subclasses,
 confirming the results by Ghisellini et al. (2014) and Chen (2018).
 2. For the 66 blazars with available Doppler factors,
    a strong correlation between the broad line emission and the Doppler factor is found.
    The correlation between the $\gamma$-ray and the broad line emission exists after the
   Doppler factor effect is removed.
    Similar results for 1 and 2 also obtained between radio and broad-line emission.
3. Our analysis suggests  a robust connection between the accretion process and the jet.}
\end{abstract}
\keywords{BL Lacertae objects: general - galaxies: active - galaxies: jets - quasars: general}

\maketitle

\section{introduction}
Active galactic nuclei (AGNs) have some special observation properties,
such as core-dominated non-thermal continuum,
 superluminal motions,
 high and variable luminosity,
 high and variable polarization,
 $\gamma$-ray emissions and so on
 (Fan et al. 2016, see also Lin et al. 2017; Lin \& Fan 2018).
 Their  properties are explained using a jet model with jet pointing close to the line of sight.
 As the most powerful subclass of AGNs,
 blazars can be divided into
  flat spectrum radio quasars (FSRQs) with strong emission lines and
  BL Lacertae objects (BL Lacs) with weak  emission lines or
   no emission lines at all.
   From the Fermi detected blazars,
Ghisellini  \& Tavecchio (2009) proposed that the BL Lacs and FSRQs can be separated in the
plot of $\gamma$-ray photon  index against the $\gamma$-ray luminosity
(see also
Abdo et al. 2009;
Ackermann et al. 2015;
Chen 2018).
Ghisellini, et al. (2011) also  pointed out  that  BL Lacs and
 FSRQs can be classified based on the luminosity of the broad line region
($L_{\rm BLR}$) measured in Eddington, and the dividing value is
about $L_{\rm {BLR}}/L_{\rm {Edd}} \sim 5 \times 10^{-4}$, which was confirmed by
Sbarrato et al. (2012, see also Yang et al. 2018a,b).

In the theoretical models of jet formation,
a correlation between jet and accretion process is expected,
the power converted into the kinetic power of jet is produced from a spinning black hole
which can release its rotational energy
(Blandford and Znajek 1977),
or produced from accretion process and disc
(Blandford and Payne 1982).
In order to understand these relationships,
a key way is to search for the connection between accretion process and jet,
which has been studied in the literatures
 (Celotti et al. 1997;
  Cao and Jiang 1999;
  Sbarrato et al. 2012;
  Ghisellini et al. 2014;
  Xiong and Zhang 2014;
  Cao 2018;
  Chen 2018).
The accretion disc produces radiation to ionize the surrounding clouds,
and further to form the broad emission lines,
so the accretion disc luminosity has a certain connection with the broad line region luminosity.
In this case, we can use  the broad line region luminosity ($L_{\rm BLR}$) to
measure the accretion disc luminosity ($L_{\rm d}$),
so $L_{\rm BLR}$ can be a proxy for the invisible accretion disk luminosity ($L_{\rm d}$)
(Sbarrato et al. 2012).

For a sample of 159 steep-and flat-spectrum quasars,
Serjeant et al. (1998) proposed that the radio luminosity ($L_{\rm R}$) is a good agent for the jet.
On the other hand, the $\gamma$-ray luminosity ($L_\gamma$) can represent for bolometric luminosity
 since that the $\gamma$-ray luminosity dominates the bolometric luminosity in Fermi blazars.
 (Dondi \& Ghisellini 1995;
  Fan, et al. 1999, 2017;
 Ghisellini et al. 2011;
  Sbarrato et al. 2012;
Xie, et al.  2004),
 so the $\gamma$-ray luminosity ($L_\gamma$) is an agent for the jet.

Therefore,
 $L_{\rm BLR}$ is an proxy for accretion disc luminosity while $L_{\rm R}$ and $L_\gamma$ are proxies for the jet.
In order to explore the connection between jet and accretion radiation,
a feasible way is to explore the relationship
 between radio (or $\gamma$-ray) luminosity and broad-line luminosity.
In 1999,
 Cao and Jiang collected a sample of 198 radio-loud quasars,
 estimated the total broad-line flux,
 and obtained a significant correlation between radio and broad-line emission.
 If the redshift is limited to the range $0.5< z <1.5$,
 a correlation with a significant level of $99.93\%$ was obtained between the radio and broad-line fluxes,
 while a significant level of $99.7\%$ was obtained between the radio and broad-line luminosities.
 One should note that the effect of the synchrotron self-absorption in blazars
can lead to measured radio fluxes lower than the intrinsic radio fluxes,
at least at lower frequencies, thus underestimating the radio luminosity.
 On the other hand, the $\gamma$-ray luminosity ($L_\gamma$) can represent for bolometric luminosity
 since that the $\gamma$-ray emissions dominate the bolometric luminosity in $\gamma$-ray loud blazars
 (Dondi \& Ghisellini 1995;
 Ghisellini et al. 2011, 2014;
  Sbarrato et al. 2012),
 so the $\gamma$-ray luminosity ($L_\gamma$) is a good proxy for the jet.

Therefore,
 $L_{\rm BLR}$ is a proxy for accretion disc luminosity
 while  $L_\gamma$ is a proxy
  for the jet luminosity.
In order to explore the connection between jet and accretion process,
a feasible way is to explore the relationship
 between
  $\gamma$-ray luminosity
  and broad-line luminosity.
Sbarrato et al. (2012)
 collected 78 blazars (with measured $L_{\rm BLR}$, $L_\gamma$ and black hole mass)
 detected by the Fermi/LAT and presented in SDSS (Sloan Digital Sky Survey),
 searched the logarithmic relationship between the $L_\gamma$ and $L_{\rm BLR}$,
 and found that $\rm log L_\gamma$ correlates well with $\rm log L_{\rm BLR}$.

In 2014, Ghisellini et al. complied a  large sample of $\gamma$-ray detected sources with measured broad emission lines,
and found a correlation between jet power as measured through the $\gamma$-ray luminosity, and accretion luminosity as measured by the broad emission lines, and that the jet power dominate over the disk luminosity.

Since there is a correlation between luminosity and redshift,
  then,
  even if there is no correlation exists in intrinsic luminosity-luminosity,
  a correlation still present on observed luminosity-luminosity (M\"ucke et al 1997).
Feigelson and Berg (1983) obtained the correlations on mutual bands,
and came to another conclusion that if there is no correlation between the luminosity-luminosity,
 nor is it in the flux-flux densities.
Many works have been done to study the relationship between the $\gamma$-ray emission
and other monochromatic emission of AGNs
 (Stecker et al. 1993;
  Padovani et al. 1993;
  Salamon and Stecker 1994;
  Fan et al. 1999; 2009; 2016).
 Some of the monochromatic emissions in blazars are strongly beamed.
  When the jet  direction is close to the line of
   sight at rest frame,
 the luminosity will be enhanced by the beaming effect,  which presents as a beaming factor ( or a Doppler factor).
 Therefore the luminosity-luminosity correlations will also be influenced by
 the correlations between the luminosity  and the beaming factor.
 In this sense, we should consider the effect of the beaming effect  when we investigate
 the luminosity-luminosity correlations.

In the present paper,
we compiled a large sample of 202 Fermi blazars
  with available redshift, radio, $\gamma$-ray, and broad-line emissions
 to study the correlation between jet and accretion disk process.
The paper is arranged as follows:
 in section 2, we will give the sample and some results;
 In section 3, some discussions and conclusions are given.
 The cosmological parameters $H_0=70 \rm km s^{-1}Mpc^{-1}$, $\Omega_{\rm m}=0.27$ and  $\Omega_{\Lambda}=0.73$
have been adopted in this work.

\section{Samples and Results}
\subsection{Samples}

In order to investigate the connection between the accretion process and jet,
 we collected the broad-line data for Fermi blazars from the  literatures
  (Celotti et al. 1997;
   Cao and Jiang 1999;
   Sbarrato et al. 2012;
   Shaw et al. 2012;
   Xie et al. 2012;
   Xiong and Zhang 2014;
   Xue et al. 2016),
and the radio and $\gamma$-ray data from Fan et al. (2016).
 The total broad-line luminosity can be calculated by the total observed luminosities from all broad lines (Celotti et al. 1997),
for some original data with flux density but no luminosity of broad-line emission,
the broad-line flux density ($f_{BLR}$) are converted into luminosity ($L_{BLR}$):
 $ L_{BLR}=4\pi \nu d_L^2 f_{BLR}$,
considering K-correction for broad-line flux density:
 $ L_{BLR}=(4\pi \nu d_L^2 f_{BLR})/(1+z)$,
where $\nu $ is the frequency,
      $d_L$ is the luminosity distance,
      $z$ is the redshift.

Therefore, we obtained a sample of 202 blazars with available monochromatic radio,
$\gamma$-ray emissions and  broad-line data,
 and listed them in Table 1.
 In Table 1,
 col. 1 gives the name of the source;
 col. 2 the redshift;
 col. 3 the classification,

    F stands for FSRQs,
    B for BL Lacs,
    U for unclassified sources;
    FL stands for LSP-FSRQs,
    FI for ISP-FSRQs,
    IB for ISP-BLs,
    LB for LSP-BLs,
    HB for HSP-BLs,
    UI for ISP-unclassified sources,
    UL for LSP-unclassified sources;
 col. 4 the logarithm of radio luminosity ( $\rm log L_{R}$ )  in units of erg/s;
 col. 5 the uncertainty for  $\rm log L_{R}$;
 col. 6 the logarithm of the $\gamma$-ray luminosity ( $\rm log L_{\gamma}$ )  in units of erg/s;
 col. 7 the uncertainty for $\rm log L_{\gamma}$;
 col. 8 the references for data in col. 4 - col. 7;
 col. 9 the logarithm of the broad line luminosity ( $\rm log L_{BLR}$ )  in units of erg/s;
 col. 10 references for col. 9,
          C97: Celotti et al.(1997),
          C99: Cao et al.(1999),
          S12: Sbarrato et al.(2012),
          X12: Xie et al.(2012),
          X14: Xiong et al.(2014),
          X16: Xue et al.(2016);
col. 11 Doppler factor, $\delta$;
 col. 12 references for Doppler factor,
          F09: Fan et al. (2009),
          H09: Hovatta et al. (2000),
          LV99: L\"ahteenim\"aki \& Valtaoja (1999),
           L17: Liodakis et al. (2017),
          S10: Savolainen et al. (2010).

For the 202 blazars,
 165 are FSRQs,
 35 are BL Lacs, and
  2 are unclassified blazars.
 The redshift ($z$) is in a range from $0.031$ to $3.104$,
 the radio luminosity ($\rm log L_R$) is  from $40.18(erg/s)$ to $44.71(erg/s)$,
 $\gamma$-ray luminosity ($\rm log L_{\gamma}$) is  from $43.39(erg/s)$ to $48.01(erg/s)$,
 and the broad-line luminosity ($\rm log L_{BLR}$) is  from $41.70(erg/s)$ to $46.65(erg/s)$.
 We also obtained Doppler factors for 66 sources from the papers by
  Fan et al. (2009),
  Hovatta et al. (2009),
  Lahteenimaki and Valtaoja (1999),
  Liodakis et al. (2017), and
  Savolainen et al. (2010).

\subsection{Results}
For the relevant data in Table1,
 a linear least regression is applied to the luminosity-luminosity correlation, following results are obtained
$$ \rm log L_{\gamma}  = (0.79 \pm 0.05)log L_{BLR} + (11.21 \pm 2.30)$$
 with a correlation coefficient $r = 0.73$ and a chance probability $p = 2.43\times10^{-35}$,  and
$$ \rm log L_R = (0.78 \pm 0.05)log L_{BLR}  + (8.57 \pm 2.10)$$
  with $r = 0.76$ and $p = 4.52\times10^{-39}$ for the 202 blazars.

For the subclasses, we have
$$ \rm log L_{\gamma} = (0.72 \pm 0.08)log L_{BLR} +  (14.28 \pm 3.54)$$
 with $r = 0.58$ and $p = 4.25\times10^{-16}$, and
 $$ \rm log L_R  = (0.79 \pm 0.07)log L_{BLR} +  (8.33 \pm 3.02)$$
 with $r = 0.67$ and $p = 4.39\times10^{-23}$
for the 165 FSRQs; and
$$ \rm log L_{\gamma} = (1.07 \pm 0.12) logL_{BLR}  - (1.23 \pm 5.10)$$
 with $r = 0.84$ and $p = 9.04\times10^{-11}$, and
$$ \rm log L_R = (1.06 \pm 0.14)log L_{BLR} -  (3.25 \pm 5.91)$$
 with $r = 0.79$ and $p = 4.51\times10^{-9}$
for the 35 BL Lacs.
They are all shown Fig.1 and Fig.2.
%
%

\begin{figure}
     \centering
 \subfloat[Fig.1: Plot of radio luminosity against emission line luminosity. From the top to the bottom is for 202 Blazars, 165 FSRQs and 35 BL Lacs.]{%
     \includegraphics[width=.30\textwidth]{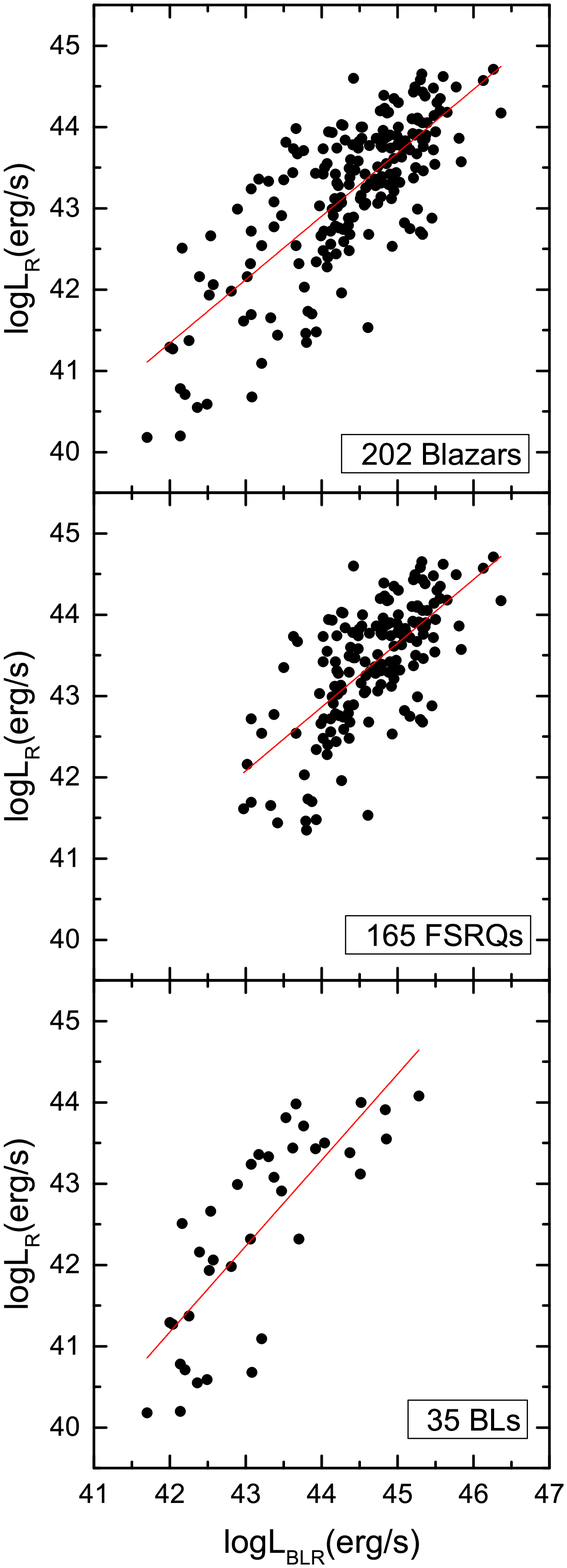}}\hfill
     \subfloat[Fig.2: Plot of $\gamma$-ray luminosity against emission line luminosity. From the top to the bottom is for 202 Blazars, 165 FSRQs and 35 BL Lacs.]{%
     \includegraphics[width=.30\textwidth]{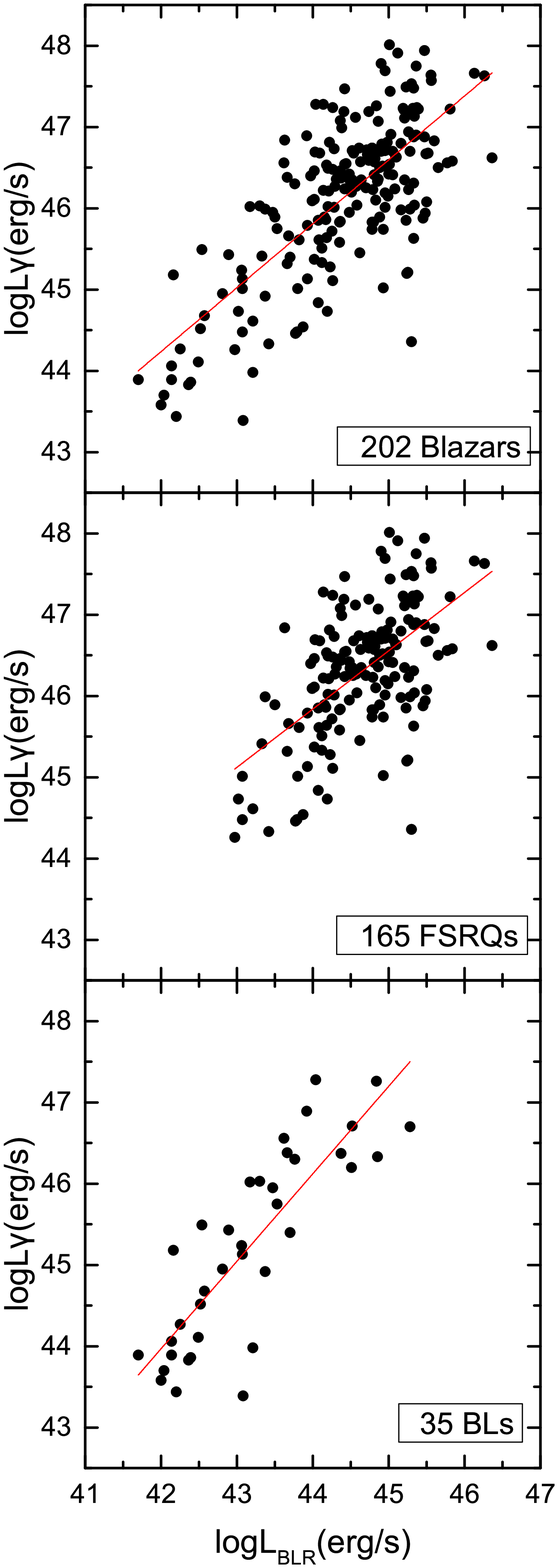}}\hfill
     \subfloat[Fig.3: Correlation between monochromatic luminosity ($\rm log L$) and Doppler factor ($\rm log \delta$). From the top to the bottom is for radio, $\gamma$-ray and broad-line luminosity against Doppler factor.]{%
     \includegraphics[width=.30\textwidth]{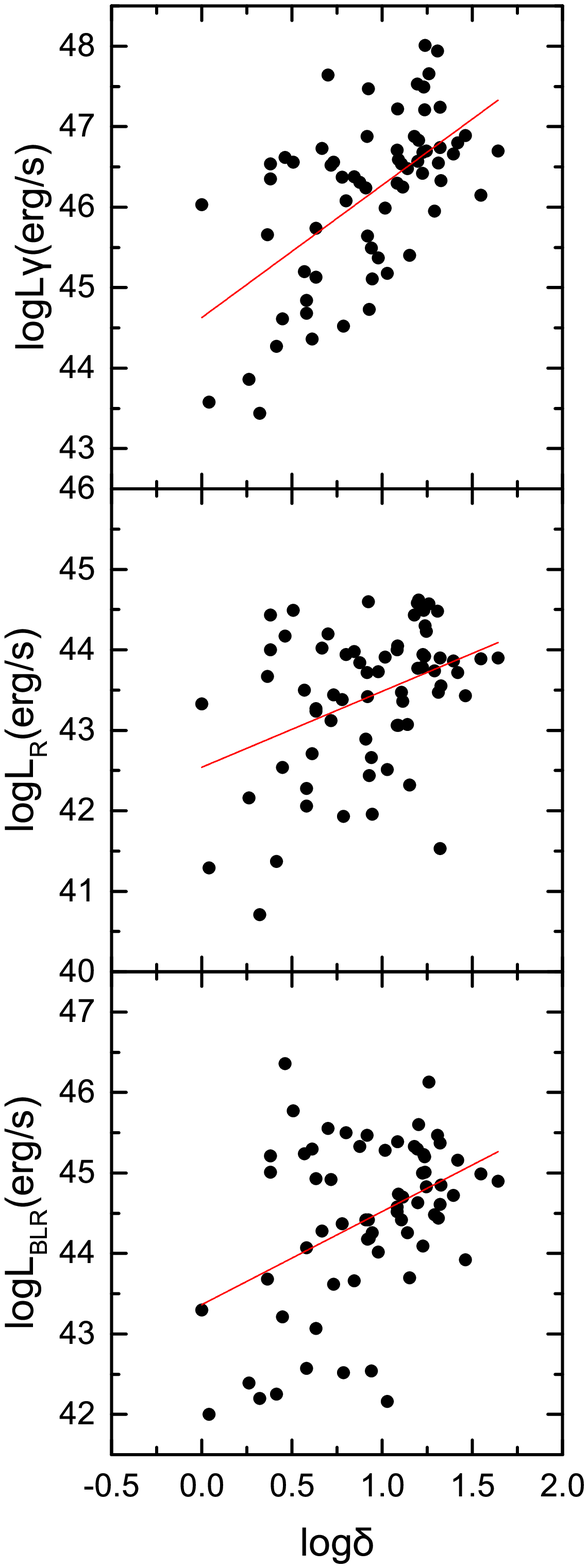}}
 \end{figure}

\section{Discussions and Conclusions}

 In the theoretical models of jet formation,
if the squared magnetic field is proportional to the accretion rate,
a correlation between jet and accretion process is expected (Ghisellini et al. 2014),
the power converted into the kinetic power of jet is produced from a spinning black hole
which can release its rotational energy
(Blandford and Znajek 1977),
or produced from accretion process and disc
(Blandford and Payne 1982).

The broad-line region is photoionized by radiation from the disc,
so the broad-line emission can be taken as a proxy of
the accretion power of the source (Celotti et al. 1997),
and can be expressed as
$L_{\rm disk} = L_{\rm BLR}/\phi$,
where $\phi \sim 0.1$
(Ghisellini et al. 2014).
For a simple one-zone leptonic model,
the  power that the jet spent in producing the non-thermal radiation is
$P_{\rm rad} = 2f{\frac{L_{\rm jet}^{\rm bol}}{\Gamma^2}},$
where
$L_{\rm jet}^{\rm bol}$ is bolometric jet luminosity,
$\Gamma$  is the bulk Lorentz factor of the outflowing plasma,
the factor 2 accounts for two jets, and
$f$ is a factor of order unity.
The power in radiation $P_{\rm rad}$ is believed to be about 10\% of the jet power
$P_{\rm jet}$, namely $P_{\rm jet} = 10 P_{\rm rad}$
 (see Ghisellini et al. 2014 for detail).
So, we can use the bolometric jet luminosity as
the proxy of the jet power and
the broad emission line luminosity as the proxy of the accretion power.

Based on the detections by EGRET and Fermi/LAT,
a series of studies have shown that there is
 a strong correlation between the $\gamma$-ray and radio emission
 (Dondi and Ghisellini 1995;
  Fan et al. 1998;
  Huang et al. 1999;
  Cheng et al. 2000;
  Abdo et al. 2009;
  Giroletti et al. 2010;
  Nieppola et al. 2011;
  Fan et al. 2016;
  Fallah et al. 2017).
Those correlations suggest that the $\gamma$-ray emission has a beaming effect.
We also found that the $\gamma$-ray luminosity and other monochromatic luminosities are strongly  correlated with Doppler factor
(Fan et al. 2017).
The beaming factor (Doppler factor) is an important parameter for blazars,
but it is difficult to determine.
Doppler factors can be obtained from a synchrotron self-Compton mechanism
(Ghisellini et al. 1993),
 from the radio variability (
Lahteennimaki and Valtaoja 1999;
Hovatta et al. 2009;
Savolainen et al. 2010;
Fan et al. 2009;
 Liodakis et al.2017), or
 by model fitting the SED of the sources
 (Ghisellini et al. 1998;
 Zhang et al. 2012, and references therein).

People think that the broad-line emission is taken as a good proxy for accretion disc.
However, the radio emission is believed from the beamed jet,
and it is also found that the $\gamma$-ray dominates the bolometric luminosity in Feremi blazars (Ghisellini et al. 2014).
In this sense,
the radio and $\gamma$-ray emission are good proxies for the jet emission while the broad-line emission is a good proxy for accretion process.
The correlation between radio and broad-line emission was studied in some literatures.
Many authors think that the correlation between the beamed emission and the broad-line emission
 is due to the close link between the relativistic jet and accretion disk (
 Celotti et al. 1997;
 Cao and Jiang 1999;
 Sbarrato et al. 2012;
 Xiong and Zhang 2014).
Cao and Jiang (1999) found a correlation between radio and broad-line flux densities for 198 radio-loud quasars.
Sbarrato et al. (2012) obtained a strong correlation between the $\gamma$-ray and broad-line luminosity.
Fan (2000) also investigated such a relationship using the EGRET data and broad-line emission.

In our present work, we compiled a large sample with Fermi/LAT detections.
Our results show a close correlation for $\rm log L_{\gamma}$-$\rm log L_{BLR}$ with
a coefficient $ r = 0.73$ and a chance probability $ p = 2.43 \times10^{-35}$,
and also a close correlation for $\rm log L_R$-$\rm log L_{BLR}$
with $ r = 0.76$ and $ p = 4.52\times10^{-39}$ for the whole sample.
Our result is consistent with those by Sbarrato et al. (2012),
Ghisellini et al (2014),and Chen (2018) for the $\gamma$-ray and broad-line emissions.
However, the sample from
Cao and Jiang (1999) was restricted to the radio flux at $5 GHz$ and the total broad-line flux.


\subsection{Redshift Effect}

Redshift effect is an important factor influencing  the  luminosity-luminosity correlation.
Considering the cross-correlation in luminosity-luminosity, it is necessary to remove redshift effect. It can be done using a formula:
$r_{ij,z} = \frac{r_{ij}-r_{iz}r_{jz}}{\sqrt{(1-r_{iz} ^2)(1-r_{jz} ^2)}}$,
here, $r_{ij}$ is the correlation coefficient in luminosity-luminosity,
$r_{iz}$ (or $r_{jz}$) is the correlation coefficient between redshift and luminosity and
$r_{ij,z}$ is the correlation coefficient in luminosity-luminosity after removing the redshift effect.

From our sample, we have got the  correlation coefficients and listed them in Table2  for the whole and subclasses,
 the coefficients after removing the redshift effect are:
$r_{L_{\gamma}L_{\rm B},z} = \frac{r_{L_{\rm R}L_{\rm B}}-r_{L_{\rm Rz}}r_{L_{\rm Bz}}}{\sqrt{(1-r_{L_{\rm Rz}} ^2)(1-r_{L_{\rm Bz}} ^2)}}
= 0.28$ with $p = 5.77 \times10^{-5}$ for $\rm log L_{\gamma}$- $\rm log L_{BLR}$, and
$r_{L_{\rm R}L_{\rm B},z}= 0.45$
with  $p = 1.33\times10^{-10}$  for $\rm log L_R$-$\rm log L_{BLR}$
for the whole sample.
For  FSRQ subclass, we have
$r_{L_{\gamma}L_{\rm B},z}$ = 0.25 with $p = 1.5\times 10^{-3}$ and $r_{L_{\rm R}L_{\rm B},z}$ = 0.49 with $p = 3.36\times 10^{-10}$
for the 165 FSRQs.
For  BL Lac subclass, we obtained
$r_{L_{\gamma}L_{\rm B},z}$ = 0.55 with  $p = 1.1\times 10^{-3}$ and
$r_{L_{\rm R}L_{\rm B},z}$  = 0.47 with $p = 5.9\times 10^{-3}$
for the 35 BL Lacs.

So, after removing the redshift effect, we can see that there are still significant correlations in $\rm log L_{\gamma}$-$\rm log L_{BLR}$ with $p = 5.77 \times10^{-5}$  and in $\rm log L_R$-$\rm log L_{BLR}$ with $p = 1.33\times10^{-10}$
 for the whole sample.
The correlations also exist for the 165 FSRQs and the 35
BL Lacs.

\subsection{Beaming Effect}

Beaming effect is important for blazars,
and included in the explanations of their observation properties.
The emission in different energy bands is produced by different mechanisms
with the radio emissions being from synchrotron emission while the high
energetic $\gamma$-rays from a synchrotron-self Compton or external Compton.
Therefore, the emission at different band has a different dependence on the Doppler factor.
In the correlation analysis of $\gamma$-ray and other low energy bands,
the $\gamma$-ray vs radio relation is found to be very strong in many literatures
(Dondi \& Ghisellini 1995;
Fan et al. 1998, 2016, 2017;
Giroletti, et al. 2010;
Huang et al. 1999;
Zhang, et al. 2000).
The strong $\gamma$-ray and radio correlation is perhaps  from the fact that  $\gamma$-ray is from an SSC mechanism or
that the $\gamma$-ray and radio emission are strongly beamed with the same Doppler factor
(Fan et al. 2009).
So, the Doppler factor estimated from the radio variability is adopted in the discussions of
beaming effect in the $\gamma$-rays.

 In this work,
radio variability Doppler factors are available for 66 sources,
and  it is interesting to find that there is a correlation between
    the broad line  luminosity
    and Doppler factor,
   see Fig.3.
Therefore, it is necessary that the Doppler factor dependence of the monochromatic luminosity and
 broad-line luminosity should be considered when we investigate the luminosity correlations.

For the 66 sources, we have
$r = 0.73$ and $p = 4.60\times10^{-12}$ for $\rm log L_{\gamma}$-$\rm log L_{BLR}$,
$r = 0.75$ and $p = 3.89 \times10^{-13}$ for $\rm log L_R$-$\rm log L_{BLR}$.
The correlation coefficients are $r = 0.59 (p= 2.01\times10^{-7}), 0.39 (p= 1.24\times10^{-3}   )$ and $0.42 (p= 5.15\times10^{-4})$ for the correlations
 between $\gamma$-ray luminosity and Doppler factor,
 between the radio luminosity and Doppler factor,
 and between broad-line luminosity and Doppler factor.
 The observed radio and $\gamma$-ray luminosities are boosted
in the jet, therefore they are correlated with Doppler factor.
For the correlation between broad-line emission and Doppler factor,
It is possible that some broad-line region  enter the jet so that
the emissions are boosted or the radiation from broad-line emission
is produced by the plasma bubbles in relativistic motion,
which show a hint that the broad-line emission clouds
have relativistic motion in the direction of sight line,
then there is an effect of jet on the broad-line clouds.
So, it is not difficult to understand that broad-line vs Doppler factor correlation.

 The correlation coefficients between the  monochromatic luminosity and broad-line luminosity after removing the Doppler factors are:
$r_{L_\gamma L_{\rm B},\delta} = 0.66$ with $p = 9.86\times10^{-8}$ for $\rm log L_{\gamma}$- $\rm logL_{BLR}$,
and $r_{L_{\rm R}L_{\rm B},\delta} = 0.70$ with $p = 1.12 \times10^{-8}$  for $\rm log L_R$-$\rm log L_{BLR}$.
We can see clearly that  the correlation between the $\gamma$-ray luminosity (or radio luminosity) and broad-line luminosity  exists even after the Doppler factor effect is removed.

Therefore, there is really a connection between the accretion process and jet even when the redshift and Doppler boosting effects are removed.
In this sense, the connection between the accretion process and jet is robust.

%
%


\subsection{Conclusions}

In this work, we compiled a sample of 202 Fermi detected blazars with available radio and emission line data.
Out of them,  66 sources have Doppler factors.
The correlations between the $\gamma$-ray luminosity (radio luminosity) and  broad-line luminosity
are discussed for the whole sample and the subclasses.
We also considered both redshift effect and beaming effect in our discussions.
Following conclusions are reached:

1) Strong correlations between the $\gamma$-ray (and radio) luminosity  and broad-line luminosity are obtained for the whole and the subclass (FSRQs and BL Lacs) samples. Those correlations exist for the whole sample and FSRQs/BL Lac sub-samples when redshift effect is removed.

2) The broad-line luminosity, the $\gamma$-ray, and the radio luminosities are all correlated with Doppler factor.

3) The correlation between the $\gamma$-ray (and radio) luminosity  and broad-line luminosity
exists after the beaming effect/(redshift effect) is removed.

4) There is a real connection between the accretion process and jet.

\begin{acknowledgements}
This work is partially supported by the National Natural Science
Foundation of  China (NSFC 11733001, NSFC U1531245, NSFC 10633010, NSFC 11173009,NSFC 11403006),  Natural Science Foundation of Guangdong Province
(2017A030313011), supports for Astrophysics  Key Subjects of Guangdong Province and Guangzhou City, and Science and Technology Program of Guangzhou (201707010401).
 JHFan is a visiting professor in the University of Padova.

\end{acknowledgements}

\begin{table}[H]
\begin{minipage}{126mm}
\caption{Fermi blazars Sample}
\label{Table1}
\begin{tabular}{@{}lccccccccccc}
\hline
   Source name &
   z  &
   class  &
    $\rm log L_R$  &
  $\Delta \rm log L_R$  &
  $\rm log L_{\gamma}$  &
  $\Delta \rm log L_{\gamma}$  &
  Ref   &
  $\rm log L_{BLR}$  &
  Ref   &
  $\delta$  &
   Ref   \\
\hline
 (1)        & (2)       & (3)       & (4)               & (5)       & (6)       & (7)               & (8)       & (9) &
 (10)        & (11)       & (12)       \\
\hline
J0017.6-0512	&	0.227	&	FI	&	41.46	&	0.02	&	44.48	&	0.05	&	F16	&	43.79	&	X14	&		&		\\
J0023.5+4454	&	1.062	&	FL	&	42.74	&	0.01	&	46.01	&	0.07	&	F16	&	44.28	&	X14	&		&		\\
J0024.4+0350	&	0.545	&	FL	&	41.35	&	0.02	&	45.01	&	0.09	&	F16	&	43.8	&	X14	&		&		\\
J0043.8+3425	&	0.966	&	FI	&	42.48	&	0.01	&	46.11	&	0.03	&	F16	&	44.02	&	X14	&		&		\\
J0046.7-8419	&	1.032	&	FL	&	43.29	&	0.04	&	45.89	&	0.11	&	F16	&	44.88	&	X14	&		&		\\
J0048.0+2236	&	1.161	&	FL	&	42.59	&	0.01	&	46.27	&	0.05	&	F16	&	44.29	&	X14	&		&		\\
J0050.4-0449	&	0.92	&	FL	&	42.88	&	0.01	&	45.83	&	0.07	&	F16	&	44.35	&	X14	&		&		\\
J0058.3+3315	&	1.369	&	FI	&	43.02	&	0.01	&	46.21	&	0.07	&	F16	&	44.21	&	X14	&		&		\\
J0105.1-2415	&	1.747	&	FL	&	43.38	&	0.02	&	46.71	&	0.06	&	F16	&	44.95	&	X14	&		&		\\
J0108.7+0134	&	2.099	&	FI	&	44.57	&	0.01	&	47.66	&	0.02	&	F16	&	46.13	&	X14	&	18.2	&	S10	\\
J0137.0+4752	&	0.859	&	FL	&	43.47	&	0.01	&	46.55	&	0.02	&	F16	&	44.44	&	X14	&	20.5	&	S10	\\
J0137.6-2430	&	0.838	&	FI	&	43.46	&	0.02	&	46.04	&	0.04	&	F16	&	45.34	&	X14	&		&		\\
J0208.6+3522	&	0.318	&	HB	&	40.2	&	0.05	&	43.89	&	0.13	&	F16	&	42.14	&	W02	&		&		\\
J0210.7-5101	&	1.003	&	UI	&	44.08	&	0.04	&	46.7	&	0.02	&	F16	&	45.28	&	Sb12	&		&		\\
J0217.1-0833	&	0.607	&	FL	&	42.72	&	0.01	&	45.01	&	0.1	&	F16	&	43.07	&	X14	&		&		\\
J0217.5+7349	&	2.367	&	FL	&	44.6	&	0.01	&	47.47	&	0.02	&	F16	&	44.42	&	C99	&	8.4	&	S10	\\
J0237.9+2848	&	1.213	&	FL	&	44.05	&	0.01	&	47.22	&	0.01	&	F16	&	45.39	&	Sh12	&	12.2	&	L17	\\
J0238.6+1636	&	0.94	&	LB	&	43.43	&	0.01	&	46.89	&	0.01	&	F16	&	43.92	&	X14	&	29	&	L17	\\
J0245.4+2410	&	2.243	&	FI	&	43.76	&	0.04	&	47.13	&	0.06	&	F16	&	45.34	&	X14	&		&		\\
J0259.5+0746	&	0.893	&	FI	&	43.35	&	0.01	&	45.89	&	0.05	&	F16	&	43.5	&	X14	&		&		\\
J0303.7-6211	&	1.351	&	FL	&	44.18	&	0.04	&	46.5	&	0.04	&	F16	&	45.65	&	X14	&		&		\\
J0309.9-6057	&	1.48	&	FL	&	44.18	&	0.04	&	46.5	&	0.03	&	F16	&	44.88	&	X14	&		&		\\
J0325.5+2223	&	2.066	&	FL	&	43.86	&	0.01	&	47.22	&	0.04	&	F16	&	45.81	&	X14	&		&		\\
J0336.5+3210	&	1.259	&	FL	&	44.17	&	0.01	&	46.62	&	0.07	&	F16	&	46.36	&	C97	&	2.9	&	L17	\\
J0339.5-0146	&	0.85	&	FL	&	43.78	&	0.01	&	46.42	&	0.02	&	F16	&	45	&	X14	&	16.7	&	L17	\\
.....         &   ....   &   ...  &   ....   &    ...   &   ...    &   ...     &   ... &   ...    &   ...  &  ...   &  ... \\
.....         &   ....   &   ...  &   ....   &    ...   &   ...    &   ...     &   ... &   ...    &   ...  &  ...   &  ... \\
.....         &   ....   &   ...  &   ....   &    ...   &   ...    &   ...     &   ... &   ...    &   ...  &  ...   &  ... \\
.....         &   ....   &   ...  &   ....   &    ...   &   ...    &   ...     &   ... &   ...    &   ...  &  ...   &  ... \\
\hline
\end{tabular}
\end{minipage}
\end{table}

\begin{table}[H]
\begin{minipage}{126mm}
\caption{Correlation Analysis Results}
\label{Table2}
\begin{tabular}{@{}lccccccccc}
\hline
  Class &
  N &
  $r_{RB}$  &
  $p_{RB}$  &
  $r_{Rz}$  &
  $p_{Rz}$  &
  $r_{Bz}$  &
  $p_{Bz}$  &
  $r_{RB,z}$  &
  $p_{RB,z}$\\
\hline
 (1)        & (2)       & (3)       & (4)               & (5)       & (6)       & (7)               & (8)       & (9) & (10)      \\
\hline
Blazars  & 202	&	0.76	&	$4.52\times10^{-39}$	&	0.77	&	$4.23\times10^{-41}$	&	0.73	&	$1.06\times10^{-34}$	&	0.45	&	 $1.33\times10^{-10}$	\\
FSRQs	& 165  &	0.67	&	$4.39\times10^{-23}$	&	0.67	&	$1.17\times10^{-22}$	&	0.55	&	$1.12\times10^{-14}$	&	0.49	&	 $3.36\times10^{-10}$	\\
BL	&  35  &	0.79	&	$4.51\times10^{-9}$	&	0.81	&	$1.95\times10^{-9}$	&	0.77	&	$3.41\times10^{-8}$	&	0.47	&	0.59\%\\	
\hline
Class & N & $r_{\gamma B}$ & $p_{\gamma B}$ & $r_{\gamma z}$ & $p_{\gamma z}$ & $r_{Bz}$ & $p_{Bz}$ & $r_{\gamma B,z}$ & $p_{\gamma B,z}$ \\ \hline
Blazars	& 202 &	0.73	&	$2.43\times10^{-35}$	&	0.88	&	$\sim 0$     	&	0.73	&	$1.06\times10^{-34}$	&	0.28	&	$5.77\times10^{-5}$	\\
FSRQs	& 165  &	0.58	&	$4.25\times10^{-16}$	&	0.84	&	$4.93\times10^{-46}$	&	0.55	&	$1.12\times10^{-14}$	&	0.25	&	 0.15\%	\\
BL   	& 35  &	0.84	&	$9.04\times10^{-11}$	    &	0.85   &	$2.03\times10^{-11}$	&	0.77	&	$3.41\times10^{-8}$ 	&	0.55	&	  0.11\%\\
\hline
\label{Table2}
\end{tabular}
\end{minipage}
\end{table}

\end{document}